\documentclass[12pt]{article}
\usepackage{amsmath,amsfonts}
\makeatletter \@addtoreset{equation}{section}

\usepackage{cite}
\usepackage{bm}
\usepackage{dcolumn}
\usepackage{color}
\makeatother
\def\one{{\hbox{ 1\kern-.8mm l}}}
\newcommand{\Dslash}{\not{\hbox{\kern-4pt $D$}}}
\newcommand{\pdslash}{\not{\hbox{\kern-2pt $\partial$}}}
\newcommand{\be}{\begin{equation}}
\newcommand{\bea}{\begin{eqnarray}}
\newcommand{\eea}{\end{eqnarray}}
\newcommand{\ba}{\begin{array}}
\newcommand{\ea}{\end{array}}
\newcommand{\ee}{\end{equation}}

\begin{document}

\begin{titlepage}
\vspace{10mm}
\begin{flushright}
 IPM/P-2014/046 \\
CERN-PH-TH-2014-184\\
\end{flushright}

\vspace*{20mm}
\begin{center}
{\Large {\bf On 3D  Minimal Massive Gravity  }\\
}

\vspace*{15mm}
\vspace*{1mm}
{Mohsen Alishahiha${}^a$, 	Mohammad M. Qaemmaqami${}^b$, Ali Naseh${}^b$ and
Ahmad Shirzad${}^{c,b}$ }

 \vspace*{1cm}

{\it ${}^a$ School of Physics, Institute for Research in Fundamental Sciences (IPM)\\
P.O. Box 19395-5531, Tehran, Iran \\ and \\
Theory Group, Physics Department, CERN, CH-1211 Geneva 23,
Switzerland \\
%${}^b$ Department of Physics, Sharif University of Technology,\\
%P.O. Box 11365-9161, Tehran, Iran\\
${}^b $ School of Particles and Accelerators\\Institute for Research in Fundamental Sciences (IPM)\\
P.O. Box 19395-5531, Tehran, Iran \\ 
${}^c $ Department of Physics, Isfahan University of Technology\\
P.O.Box 84156-83111, Isfahan, IRAN
}

 \vspace*{0.5cm}
{E-mails: {\tt \{alishah, m.qaemmaqami, naseh, shirzad\} @ipm.ir, }}%

\vspace*{2cm}
%%\maketitle
\end{center}

\begin{abstract}
We study linearized equations of motion of the newly proposed three dimensional gravity, known as
minimal massive gravity, using its metric formulation. 
By making use of a redefinition of the parameters of the model, we  observe that the resulting linearized equations 
are exactly the same as that of TMG . In 
particular the model admits logarithmic modes at  critical points.
We also study several vacuum solutions of the model,  specially  at a certain 
limit where the contribution of Chern-Simons term vanishes.
\end{abstract}

\end{titlepage}

\section{Introduction}

Minimal Massive Gravity (MMG) \cite{Bergshoeff:2014pca} is a three dimensional massive gravity which has the same  
minimal local structure as  Topologically Massive Gravity (TMG)\cite{Deser:1981wh}.   Working within  the ``Chern-Simons-like'' formulation of massive gravity \cite{Hohm:2012vh} one may 
define  the  action of
the MMG model from that of TMG by adding an extra term. 
Adding this extra term would open up the possibility to get a consistent ghost free and non-tachyonic
three dimensional theory. Nevertheless its linearization about a  flat or AdS vacuum still has a single massive mode.

To explore this point better, let us first consider the TMG model 
in the Chern-Simons like formulation whose  Lagrangian  3-form  may be written as follows\cite{Bergshoeff:2014pca}
\be\label{act1}
L_{\rm TMG}=-\sigma e\cdot R+\frac{\Lambda_0}{6} \; e\cdot e\times e+h\cdot T(\omega)+\frac{1}{2\mu}
\left(\omega \cdot d\omega+\frac{1}{3}\omega\cdot \omega\times \omega\right),
\ee
where $\sigma$ is a sign, $\Lambda_0$ is a primary cosmological constant, $e$ and $\omega$ are  dreibein  and dualised spin-connection, respectively. $\mu$ is a  mass parameter of the model.

In terms of these variables 
the Lorentz covariant torsion and curvature 2-forms  are given by 
\be
T(\omega)=de+\omega\times e,\;\;\;\;\;\;\;\;\;\;\;R(\omega)=d\omega+\frac{1}{2}
\omega\times \omega.
\ee
In the action \eqref{act1}  the auxiliary field $h$ may be thought of as a  Lagrange multiplier that imposes the zero torsion constraint. 

It is then straightforward to find  equations of motion of the above action by taking the variation 
with respect to the fields $h, e$ and $\omega$
\be
 T(\omega)  =0,\;\;\;
 \sigma R(\omega)-\frac{\Lambda_{0}}{2} e\times e -D(\omega)h=0,\;\;\; R(\omega)+\mu e\times h +\sigma\mu T(\omega) = 0.
\ee
For generic values of $\Lambda_0$ and $\mu$, these  equations admit several solutions  including 
AdS, BTZ  and warped AdS black holes  solutions\cite{{Moussa:2003fc},{Bouchareb:2007yx},{Anninos:2008fx},{Moussa:2008sj}}. 
It is, however, known that this model suffers from the fact that the energy of graviton and the 
mass of BTZ black holes cannot be positive at the same time; tuning the parameters to get a 
positive energy for the graviton would result in BTZ solutions with negative mass, and vice
 versa, {\it i.e} a positive mass BTZ black hole gives a negative energy for graviton. 
Alternatively, if one thinks of the model as one which 
provides a gravitational dual for a two dimensional conformal field theory (CFT), the corresponding 
CFT would have negative central charge whenever the propagating spin-2 mode in the bulk has positive energy.

In order to have a well defined unitary model, it was proposed in \cite{Li:2008dq} that the parameters of 
the TMG may be tuned to particular values (critical points), so that the modes with negative 
mass are removed from the model leading to  a gravitational theory which could 
provide a holographic description for a chiral CFT. It was, however, observed \cite{Grumiller:2008qz} that although at 
the ``critical points'' the modes with negative mass are removed from the theory, the model
 admits new ``logarithmic'' modes which again results in a non-unitary theory\footnote{For logarithmic solution in higher dimensional gravity see, e.g.  \cite{{Alishahiha:2011yb},{Gullu:2011sj},{Bergshoeff:2011ri}, Johansson:2012fs}.}.

To circumvent the above problem, the authors of  \cite{Bergshoeff:2014pca}, proposed  
a new model, named MMG, whose Lagrangian  3-form  is
\be
L_{\rm MMG}=L_{\rm TMG}+\frac{\alpha}{2}\; e\cdot h\times h.
\ee
Then, it was  shown that in the specific range of the parameters of the model, it is possible to get 
positive central charges for the dual conformal field theory and, at the same time to have a graviton with positive energy. 

The equations of motion derived from the action of MMG are given by\cite{Bergshoeff:2014pca}
\bea\label{eq}
 T(\omega) - \alpha e\times h =0,\;\;\;\;\;\;\;\;\;\; R(\omega)+\mu e\times h +\sigma\mu T(\omega) = 0,
\eea
and
\bea\label{eq1}
 -\sigma R(\omega)+\frac{\Lambda_{0}}{2} e\times e +D(\omega)h+\frac{\alpha}{2} h\times h =0.
\eea
Defining $\Omega=\omega+\alpha h$ one can solve the equations in \eqref{eq} to find $\Omega$ and
$h$ in terms of $e$. Then plugging the results into the equation \eqref{eq1} one can find an equation for 
the metric as follows
\bea\label{e.o.m.metric}
\bar{\sigma} G_{\mu\nu} +\bar{\Lambda}_{0} g_{\mu\nu} +\frac{1}{\mu} C_{\mu\nu} +\frac{\gamma}{\mu^2}J_{\mu\nu}=0,
\eea
where the Einstein tensor $G_{\mu\nu}$,  and the Cotton tensor $C_{\mu\nu}$  are
given by 
\bea
G_{\mu\nu} = R_{\mu\nu} -\frac{1}{2}R g_{\mu\nu},\;\;\;\;\;\;\;
C_{\mu\nu} =\epsilon_{\mu}\;^{\alpha\beta} \nabla_{\alpha}(R_{\beta\nu}-\frac{1}{4}g_{\beta\nu}R)
\eea
while the curvature-squared symmetric tensor   $J_{\mu\nu}$ is defined by\footnote{In our convention, $\epsilon^{123}=\frac{1}{\sqrt{-g}}$ and $\epsilon_{123}= -\sqrt{-g}$.}
\be
J_{\mu\nu} = R_{\mu}\;^{\alpha}R_{\alpha\nu} -\frac{3}{4}R R_{\mu\nu}-\frac{1}{2}g_{\mu\nu} \left(R^{\alpha\beta}R_{\alpha\beta}-\frac{5}{8}R^{2}\right).
\ee
Here the parameters $\gamma, \bar{\sigma}$ and $\bar{\Lambda}_0$ are defined in terms of the parameters of the action as follows\footnote{Since
 our convention for Levi-Civita tensor differs from that in  \cite{Bergshoeff:2014pca}
{by} a minus sign,  {we will get} an extra  overall minus sign in our following equations.} 
 \bea\label{compatibility.relations}
\gamma = \frac{\alpha}{(1+\sigma\alpha)^2},~~\bar{\sigma} = - \left(\sigma + \alpha+\frac{\alpha^2\Lambda_{0}}{2\mu^{2}(1+\sigma\alpha)^{2}}\right),~~
\bar{\Lambda}_{0} = -\Lambda_{0} \left(1+\sigma\alpha-\frac{\alpha^{3}\Lambda_{0}}{4\mu^{2}(1+\sigma\alpha)^{2}}\right).
\eea
It is argued in \cite{Bergshoeff:2014pca}  that the  equation \eqref{e.o.m.metric}  for $\alpha\neq 1$ cannot be obtained from
any conventional deformation of the TMG action with higher order curvature terms.

It is the aim of this article to further study the MMG model. In particular we will reconsider the linearization of MMG around an AdS vacuum solution in the metric formulation of the model. 
We will see that 
the resulting equations are exactly the same as that of TMG upon a redefinition of 
the parameters of the model. This shows that TMG and MMG have locally the same 
structure, though when it 
comes to computing the conserved charges they behave differently. 
This is, indeed, the reason why MMG could provide a well defined three dimensional 
gravity with minimal degrees of freedom.
We shall also see that there is a point in the moduli space of parameters
of the model, where the linearized equations of motion reduce into that of 3 dimensional conformal gravity.

We will also study wave solutions of the MMG model, where one observes that, 
just like the TMG model,  there are  critical points at 
which the model exhibits logarithmic solutions, though unlike the TMG case, a priori one is not forced 
to be at the critical points. Having found a logarithmic solution for particular values of the parameters of the theory, it is then natural 
to propose that the MMG model could also provide a holographic description for a logarithmic CFT (LCFT) at the critical point. Following 
this idea we will also compute the new anomaly of the corresponding  LCFT.

 We shall also study other possible solutions of the MMG model, specially 
those with zero contribution from the  Chern-Simons term.
 Motivated by these solutions we will
study the model in a limit where $\mu\rightarrow \infty, \gamma\rightarrow \infty$, while keeping 
$\kappa=\frac{\gamma}{\mu^2}$ fixed.  In this limit the Chern-Simons terms drops though
the curvature-squared term survives the limit. Note that
in the Chern-Simons like formulation of MMG, this
limit corresponds to a limit in which $\mu\sim \frac{1}{\epsilon}, 1+\sigma\alpha\sim \epsilon$
for $\epsilon\rightarrow 0$.

The paper is organized as follows. In the next section we will study the linearization of MMG 
in the metric formulation where we will show that the model has several critical points at which 
it becomes either conformal, or admits logarithmic modes. 
In section three we study wave solutions of the model where we will find
logarithmic solutions at the critical points.
 In section four we will study new solutions of the MMG equations of motion specially in the limit
where the contribution of Chern-Simons term vanishes. 
Clearly the obtained solutions cannot be  solutions of TMG, nor Einstein gravity. 
In section five we will compute the new anomaly of the corresponding dual LCFT.  
The last section is devoted to conclusions.

%%%%%%%%%%%%%%%%%%%%%%%%%%%%%%%%%%%%%%%%%%%%%%%%%%%%%%%%%%%%%%%%%%

\section{Linearized analysis in metric formalism}

In this section,  using the metric formulation of MMG  given by the equation \eqref{e.o.m.metric}, we study linear excitations of metric around an AdS$_{3}$ vacuum. Denoting the vacuum metric by $\bar{g}_{\mu\nu}$,
one sets
\bea
g_{\mu\nu} = \bar{g}_{\mu\nu} + h_{\mu\nu},
\eea
where $h_{\mu\nu}$  is the small perturbation and $\bar{g}_{\mu\nu}$ satisfies the following 
equation
\bea\label{l}
\bar{R} _{\mu\nu} = -\frac{2}{l^2}\bar{g}_{\mu\nu},\;\;\;\;\;\;{\rm with}\;\;\;\;\;\;
l^2 = -\frac{1}{2\mu}\frac{1}{\bar{\Lambda}_{0}}(\mu\bar{\sigma} \pm \sqrt{\mu^{2}\bar{\sigma}^{2}-\bar{\Lambda}_{0}\gamma}).
\eea
At the linearized level, the equation of motion (\ref{e.o.m.metric}) reduces to
\footnote{Here and in what follows, the covariant derivative is defined with respect to the background metric $\bar{g}_{\mu\nu}$. }
\be\label{Lin1}
\bar{\sigma} (R_{\mu\nu}^{(1)}-\frac{1}{2}\bar{g}_{\mu\nu}R^{(1)}+\frac{3}{l^2} h_{\mu\nu})+\bar{\Lambda}_{0} h_{\mu\nu}+\frac{1}{\mu} \epsilon _{\mu}\;^{\alpha\beta}\bar{\nabla}_{\alpha}\left(
R_{\beta\nu}^{(1)}-\frac{1}{4}\bar{g}_{\beta\nu}R^{(1)}+\frac{2}{l^2} h_{\beta\nu}\right) =\frac{-\gamma}{2\mu^{2}l^2}J_{\mu\nu}^{(1)}, 
\ee
where 
\bea
J^{(1)}_{\mu\nu} =  R_{\mu\nu}^{(1)}-\frac{1}{2} \bar{g}_{\mu\nu}R^{(1)}+\frac{5}{2l^2}h_{\mu\nu}.
\eea
Moreover  the linearized Ricci tensor and Ricci scalar are given by
\bea
R_{\mu\nu}^{(1)} &=& \frac{1}{2}\left(-\bar{\nabla}^{2}h_{\mu\nu}-\bar{\nabla}_{\mu}\bar{\nabla}_{\nu}h
+\bar{\nabla}_{\mu}\bar{\nabla}_{\sigma}h^{\sigma}_{\nu}+\bar{\nabla}_{\nu}\bar{\nabla}_{\sigma}h^{\sigma}_{\mu}
+\frac{2}{l^2}\bar{g}_{\mu\nu}h-\frac{6}{l^2} h_{\mu\nu}\right),\cr\nonumber\\
R^{(1)} &=& -\bar{\nabla}^{2}h +\bar{\nabla}_{\rho}\bar{\nabla}_{\sigma}h^{\rho\sigma}+\frac{2}{l^{2}}h,
\eea
with $h \equiv \bar{g}^{\mu\nu} h_{\mu\nu}$.
Using these expressions and in  the gauge $
\bar{\nabla}^{\mu}h_{\mu\nu} = \bar{\nabla}_{\nu} h$ the equation \eqref{Lin1} reads
\bea\label{gauged.e.o.m}
(\bar{\sigma}+\frac{\gamma}{2\mu^{2}l^2})\big(-\frac{1}{2}\bar{\nabla}^{2}h_{\mu\nu}
+\frac{1}{2}\bar{\nabla}_{\mu}\bar{\nabla}_{\nu}h-\frac{3}{l^2} h_{\mu\nu}\big)
+ \big(\frac{3\bar{\sigma}}{l^2}+\bar{\Lambda}_{0}+\frac{5}{4}\frac{\gamma}{\mu^{2}l^4}\big)h_{\mu\nu}&&
\cr &&\cr
+\frac{1}{\mu}\epsilon_{\mu}\;^{\alpha\beta}\bar{\nabla}_{\alpha}\big(
-\frac{1}{2}\bar{\nabla}^{2}h_{\beta\nu}+\frac{1}{2}\bar{\nabla}_{\nu}\bar{\nabla}_{\beta}
h+\frac{1}{2l^2}\bar{g}_{\beta\nu}h-\frac{1}{l^2} h_{\beta\nu}\big) &=& 0.\nonumber\\
\eea
On the other hand from the  trace of the equation (\ref{gauged.e.o.m}) one finds 
\bea
(4\bar{\Lambda}_{0}-\frac{\gamma}{\mu^{2}l^4}) h =0, 
\eea
which for  $(4\bar{\Lambda}_{0}-\frac{\gamma}{\mu^{2}l^4}) \neq 0$,  sets the trace of  the perturbation, $h$, to zero. By making use of the linearized Bianchi identity and setting $h=0$, the equation (\ref{gauged.e.o.m}) may be recast into the following form 
\bea\label{guaged.e.o.m.m}
\bigg(\bar{\nabla}^{2}+\frac{\gamma - 4\mu^{2}l^4\bar{\Lambda}_{0}}{l^2(\gamma +2\mu^{2}l^2\bar{\sigma})}\bigg)h_{\mu\nu} +\frac{2\mu l^2}{\gamma+2\mu^{2}l^2\bar{\sigma}}\left(\bar{\nabla}^{2}+\frac{2}{l^2}\right)\epsilon_{\mu}\;^{\alpha\beta}
\;\bar{\nabla}_{\alpha}h_{\beta\nu} =0 .
\eea
Furthermore using the relation between $l$ and $\bar{\Lambda}_0$ given  in the equation \eqref{l} one arrives at
\bea\label{pre.equation}
\left(\bar{\nabla}^{2}+\frac{2}{l^2}\right) \bigg(\frac{1}{\tilde{\mu}}\epsilon_{\mu}\;^{\alpha\beta}\;
\bar{\nabla}_{\alpha}h_{\beta\nu}+h_{\mu\nu}\bigg) =0,\;\;\;\;\;\;\;\;\;\;{\rm with}\;\;\;\;\;
\tilde{\mu}=\frac{\gamma+2\mu^{2}l^2\bar{\sigma}}{2\mu l^2}.
\eea
Note that in this notation the mass of the massive graviton is 
\be\label{eqmass}
M^2=\frac{\tilde{\mu}^2 l^2-1}{l^2}.
\ee
It is interesting to note that this equation is exactly the same as that found from the linearization of TMG around an AdS vacuum  (see the equation 29 \cite{Li:2008dq}), with the replacement of $\mu$ with $\tilde{\mu}$. This, in turn, shows
that locally TMG and MMG have the same degrees of freedom as emphasised in\cite{Bergshoeff:2014pca}.
Note that for $\gamma=0$ one has $\tilde{\mu}=-\sigma \mu$.

Having realized the similarity between the linearized equation of MMG and that of TMG one can utilize the
results of \cite{Li:2008dq} to find the spectrum of the perturbation of MMG around an AdS vacuum. Indeed,
following \cite{Li:2008dq} one may rewrite the linearized equation \eqref{pre.equation} in terms of the
quadratic Casimir operators of the $SL(2,R)_L\times SL(2,R)_R$ algebra associated with the  isometry group of the AdS background 
geometry as follows \cite{Li:2008dq}
\bea
\big[2 (L^{2}+\bar{L}^{2}) + 3+\tilde{\mu}^{2}l^2\big] \big[L^{2}+\bar{L}^{2}
+{2}\big] h_{\mu\nu} = 0.
\eea
Therefore the solutions of the linearized equation can be classified by the representation of 
$SL(2,R)_L\times SL(2,R)_R$ algebra. More precisely consider  primary states $|\psi_{\mu\nu}\rangle$ with weight $(h,\bar{h})$, so that
\bea
&&L_0 |\psi_{\mu\nu}\rangle =h|\psi_{\mu\nu}\rangle,\;\;\;\;\;\;\;\;\;\;
L_1|\psi_{\mu\nu}\rangle =0,\cr&&\cr
&&\bar{L}_0 |\psi_{\mu\nu}\rangle =\bar{h}|\psi_{\mu\nu}\rangle,\;\;\;\;\;\;\;\;\;\;
\bar{L}_1|\psi_{\mu\nu}\rangle =0.
\eea
Thus, in the present case the  primary weights $(h,\bar{h})$ satisfy the following equation
\bea\label{eql}
\big[2h(h-1)+2\bar{h}(\bar{h}-1)-3-\tilde{\mu}^{2}l^2\big]\big[h(h-1)+\bar{h}(\bar{h}-1)-2\big] = 0.
\eea
Here we have used the fact that 
\bea
L^{2} \mid \psi_{\mu\nu} \rangle = -h(h-1) \mid \psi_{\mu\nu} \rangle, ~~~~~~~ \bar{L}^{2} \mid \psi_{\mu\nu} \rangle = -\bar{h}(\bar{h}-1) \mid \psi_{\mu\nu} \rangle,
\eea
Note also that from the gauge condition one has $h-\bar{h}=\pm2$. 

The equation \eqref{eql}  has two branches of solutions where either of its factors is zero 
\bea\label{Bran}
{\rm Branch\; 1}&:&(h^{(0)},\bar{h}^{(0)})=~\Bigg\{ \begin{array}{rcl}
&(\frac{3}{2} \pm \frac{1}{2}, -\frac{1}{2} \pm \frac{1}{2}) ,& ~~~~~~~~~~~h^{(0)}-\bar{h}^{(0)}=2,\\ &\\
&(-\frac{1}{2} \pm \frac{1}{2},  \frac{3}{2} \pm \frac{1}{2}),& ~~~~~~~~~~~h^{(0)}-\bar{h}^{(0)} = -2,
\end{array}\, \nonumber \\ \nonumber  &&\\ 
{\rm Branch\; 2}&:&(h^{(m)},\bar{h}^{(m)})=\Bigg\{ \begin{array}{rcl}
( \frac{3}{2} \pm \frac{1}{2} \tilde{\mu} l, -\frac{1}{2} \pm \frac{1}{2} \tilde{\mu} l ),& ~~~~~~h^{(m)}-\bar{h}^{(m)} =2,\\ &\\
(-\frac{1}{2} \pm \frac{1}{2} \tilde{\mu} l, \frac{3}{2} \pm \frac{1}{2} \tilde{\mu} l ),&~~~~~~~~h^{(m)}-\bar{h}^{(m)} =-2.
\end{array}\, 
\eea
As one observes, the spectrum is exactly the same at that of the TMG model. In particular for 
normalizable modes where one has to keep the upper signs in the above equation, the first branch
corresponds to left and right moving massless gravitons: $(2,0), (0,2)$, while from the second branch and
taking into account that the equation is invariant under $\tilde{\mu}\rightarrow -\tilde{\mu}$, one gets
normalizable modes corresponding to the  massive gravitons $ ( \frac{3}{2} - \frac{1}{2} \tilde{\mu} l, -\frac{1}{2} - \frac{1}{2} \tilde{\mu} l )$.

Although unlike the TMG model, a priori, we are not forced to fix the parameter $\tilde{\mu} l$, one could still have
critical points $\tilde{\mu}l=\pm 1$ at which the massive gravitons  degenerate with  either of left or 
right massless gravitons. In terms of the original parameters of the model in the metric formulation the critical points are given by
\bea\label{gamma.c}
\gamma = -2\mu l (\mu l\bar{\sigma} \mp 1),
\eea
while in terms of  the parameters appearing in the Chern-Simons-like formalism, using the equation \eqref{compatibility.relations}, they occur at 
\bea\label{critical.values.Drei.1}
\alpha = -2 (\sigma\pm \frac{1}{\mu l }).
\eea

Since the linearized equations of motion for MMG has locally the same form as TMG, one may wonder if the model reduces to a three dimensional conformal gravity in the limit of $\tilde{\mu}\rightarrow 0$. Indeed 
form the definition of   $\tilde{\mu}$ one observes that $\tilde{\mu}=0$ corresponds to
$\gamma= -2\mu^{2}l^{2}\bar{\sigma}$. On the other hand at this point 
the coefficients of the first and second terms in the equation  (\ref{gauged.e.o.m}) vanishe, yielding to
\be
\frac{1}{\mu}\epsilon_{\mu}\;^{\alpha\beta}\bar{\nabla}_{\alpha}\big(
-\frac{1}{2}\bar{\nabla}^{2}h_{\beta\nu}+\frac{1}{2}\bar{\nabla}_{\nu}\bar{\nabla}_{\beta}
h+\frac{1}{2l^2}\bar{g}_{\beta\nu}h-\frac{1}{l^2} h_{\beta\nu}\big) = 0,
\ee
which is just the contribution from the Cotton tensor. In other words, for this special value of $\gamma$ the theory of MMG, at the linearized level,  has locally the same content as that of 3D conformal gravity.

%%%%%%%%%%%%%%%%%%%%%%%%%%%%%%%%%%%%%%%%%%%%%%%%%%%%%%%%%%%%%%%%%%%
\section{Wave solutions}
%\subsection{Wave solution in metric formalism}

In the previous section we have seen  that  the MMG model has an AdS vacuum and small perturbations around 
this vacuum have the same structure as that of TMG. 
Since the TMG model, besides the AdS vacuum, has other vacuum solutions, it is then natural
to look for other solutions of the MMG model too. In particular one may study  wave solutions 
in this model\footnote{The  wave solutions of TMG and NMG with logarithmic profile have been 
first studied in 
\cite{{AyonBeato:2004fq},{AyonBeato:2005qq}}.}.
To proceed we consider an ansatz for a wave solution as follows
\be
g_{\mu\nu}=\bar{g}_{\mu\nu}+F k_\mu k_\nu,
\ee 
with $k_\mu$ being  a null vector with respect to the background AdS metric $\bar{g}_{\mu\nu}$ which 
could be parametrized  as 
\bea
ds^{2} = \frac{l^{2}}{r^{2}} (dr^{2}-2dudv),
\eea
where $l$ is given in equation \eqref{l} and $u,v$ are light-like coordinates. With this notation, our ansatz for the wave solution takes the
following form
\bea\label{AdS.wave}
ds^{2} = \frac{l^{2}}{r^{2}} \bigg(dr^{2}-2dudv - F(r,u)du^{2}\bigg).
\eea
Here $F(r,u)$ is an arbitrary function which can be determined by equations of motion. By substituting
this ansatz into the equations of motion (\ref{e.o.m.metric}), one finds 
\bea
\frac{1}{4\mu^{2}l^{2}r}\left[-2\mu l{r}^{2}{\frac{\partial^{3}F}{\partial{r}^{3}}}+(2{\mu}^{2}{l}^{2}\bar{\sigma}+\gamma)\bigg(r{\frac{\partial^{2}F}{\partial{r}^{2}}}-{\frac{\partial F}{\partial r}}\bigg)\right]=0.
\eea
Using the definition of $\tilde{\mu}$ given in the equation \eqref{pre.equation}, the above differential
equation may be recast into the following form
\be\label{eq3}
-{r}^{2}{\frac{\partial^{3}F}{\partial{r}^{3}}}+\tilde{\mu}l \bigg(r{\frac{\partial^{2}F}{\partial{r}^{2}}}-{\frac{\partial F}{\partial r}}\bigg) = 0,
\ee
which can be solved to give a generic AdS-wave solution with
\bea\label{sol1}
F(r,u) = F_{0}(u)+F_{2}(u)r^{2}+F_{\tilde{\mu}}(u)~r^{1+\tilde{\mu}l}.
\eea
It is then obvious that the solution  degenerates at the critical points $\tilde{\mu}l=\pm 1$, as
expected.  Indeed, just like TMG, one would expect that the MMG model exhibits logarithmic
solutions at the critical points. In fact at these points one gets
\bea\label{critical.value.gamma}
\tilde{\mu}l=-1&&({\rm or}\;\gamma= -2\mu l(\mu l\bar{\sigma}+1))~~~\rightarrow ~~~ F(r,u) =\tilde{F}_{0}(u)\log(r)+F_{0}(u)+F_{2}(u)r^{2}\nonumber \\ \\ \tilde{\mu}l=1
&&({\rm or}\;\gamma= -2\mu l(\mu l\bar{\sigma}-1))~~~\rightarrow ~~~ F(r,u) = F_{0}(u)+F_{2}(u)r^{2}+\tilde{F}_{2}(u)r^{2}\log(r).\nonumber
\eea
It is important to note that  at the first critical point where  $\tilde{\mu}l=-1$, the model admits a solution which is not asymptotically locally AdS geometry.  In other words in the context of Gauge/Gravity duality, the leading log-term corresponds to turning on an irrelevant operator in the dual field theory. Actually one would expect that the dual field theory  to be a LCFT. On the other hand at the second critical point where
$\tilde{\mu}l=1$,  the geometry is, indeed,  asymptotically locally AdS, though, comparing with 
an AdS geometry,  its field theory dual is now in a new vacuum.

We note also that at $\tilde{\mu}=0$ the solution generates a linear term in the expression of $F$
\bea\label{conformal.solution}
 F(r,u) = F_{0}(u)+F_{1}(u)r + F_{2}(u)r^{2}.
\eea  
indicating that  the corresponding gravity should come from a 3D conformal gravity, as 
anticipated in the
previous section. Note that since $k_\mu$ is a null vector, the first order perturbation given by
$h_{\mu\nu}=F k_\mu k_\nu$ is, indeed, an exact solution of the whole equations of motion.

Note also that setting the integration constants $F_0$ and $F_2$ in the solution  
\eqref{sol1} to zero and assuming $\tilde{\mu}l<-1$ one finds the three dimensional 
 Schr\"odinger solution (null solution) with dynamical scaling $z=\frac{1}{2}(1-\tilde{\mu} l)$
\be
ds^2=l^2\left(\frac{dr^2}{r^2}-\frac{2 dv du}{r^2}-\frac{du^2}{r^{2z}}\right).
\ee
Null solutions in TMG and NMG have also been studied in \cite{Anninos:2008fx} 
and \cite{Clement:2009ka}, respectively.

It is worth mentioning that one could have found the wave solutions and the corresponding 
critical points from  the Chern-Simons like formulation of MMG. Actually for our ansatz \eqref{AdS.wave}
one has
\bea
e^{1} = \frac{l}{r} dr,~~~~~~~~ e^{2} = \frac{1}{2}F(r,u)du + dv,~~~~~~~~ e^{3} =\frac{l^{2}}{r^{2}}du.
\eea
which can be used to read the dual spin-connection from the torsion-less condition
\be
\Omega_{r}^{1} = -\frac{1}{r},~~~~~~\Omega_{u}^{2} =\frac{1}{2l}(-r\frac{\partial F}{\partial r}+F),~~~~~~~
\Omega_{u}^{3} = -\frac{l}{r^{2}},~~~~~~\Omega_{v}^{2} = \frac{1}{l}.
\ee
Moreover, from the second equation in (\ref{eq}), one can find the  non-zero components of $h_{\mu}^{a}$ 
as follows
\bea
&&h_{r}^{1} = \frac{1-\alpha l^{2}\Lambda_{0}}{2\mu lr (1+\sigma\alpha)^{2}},~~~~~~h_{u}^{3} =  \frac{1-\alpha l^{2}\Lambda_{0}}{2\mu r^{2} (1+\sigma\alpha)^{2}}, ~~~~~~h_{v}^{2} =  \frac{1-\alpha l^{2}\Lambda_{0}}{2\mu l^{2}(1+\sigma\alpha)^{2}},\cr\nonumber\\\nonumber\\
&&~~~~~~h_{u}^{2} = -\frac{1}{4\mu l^{2}(1+\sigma\alpha)^{2}} \left(
-2r^{2}\frac{\partial^{2}F}{\partial r^{2}}+2r\frac{\partial F}{\partial r}+(\alpha l^{2}\Lambda_{0}-1)F\right),
\eea
where 
\bea\label{Lambda0}
\Lambda_{0} = \frac{1}{\alpha l^{2}} +\frac{2\mu^{2}(1+\sigma\alpha)^{3}}{\alpha^{3}} \left(
1 \mp \sqrt{1+\frac{\alpha^{2}}{\mu^{2}l^{2}(1+\sigma\alpha)^{2}}}~\right).
\eea
Plugging everything into the equation \eqref{eq1}, for $1+\sigma\alpha\neq 0$, one 
arrives at\footnote{Note that in order to get a well defined limit at $\alpha=0$ one needs
to choose the upper sign.} 
\bea\label{Ads.wave.Der}
-r^{2}\frac{\partial^{3}F}{\partial r^{3}}-\frac{\mu l}{\alpha}\bigg[(1+\sigma\alpha)^{2}\mp\sqrt{(1+\sigma\alpha)^{2}+\frac{\alpha^{2}}{\mu^{2}l^{2}}}~~\bigg]\bigg(r\frac{\partial^{2}F}{\partial r^{2}}-\frac{\partial F}{\partial r}\bigg)=0,
\eea
which is exactly the same as \eqref{eq3}, taking into account  the following  identity
\be\label{Id}
\tilde{\mu} =-\frac{\mu }{\alpha}\bigg[(1+\sigma\alpha)^{2}-\sqrt{(1+\sigma\alpha)^{2}+\frac{\alpha^{2}}{\mu^{2}l^{2}}}~~\bigg].
\ee

%%%%%%%%%%%%%%%%%%%%%%%%%%%%%%%%%%%%%%%%%%%%%%%%%%%%%%%%%%%%%%%%%%%%%%%%
%%%%%%%%%%%%%%%%%%%%%%%%%%%%%%%%%%%%%%%%%%%%%%%%%%%%%%%%%%%%%%%%%%%%%%%%

\section{On new  MMG vacuum solutions}

It is known that any Einstein solution is a solution of TMG and since locally MMG has the same structure
as that of TMG, it should also be a solution of MMG. However,   TMG has more solutions 
than just the Einstein solutions and therefore  it is natural to expect that
MMG should also have more solutions than Einstein or even TMG itself.

In particular one can see that for every value of $\mu l\neq 3$ in TMG, there are
vacuum solutions with $SL(2,R)\times U(1)$ isometry known as warped AdS geometries
\cite{{Moussa:2003fc},{Bouchareb:2007yx},{Anninos:2008fx},{Moussa:2008sj}} (see also\cite{Clement:2009gq} for NMG). The 
$U(1)$ factor in the isometry group could be timelike or spacelike.  Recently it has been shown \cite{Arvanitakis:2014yja} that the MMG model also admits warped AdS solutions. More precisely,
starting with the following ansatz
\bea
ds^2=\frac{l^2}{\nu^2+3}\left[-\cosh^2\rho d\tau^2+d\rho^2+\frac{4\nu^2}{\nu^2+3}(d\xi+\sinh\rho d\tau)^2\right],
\eea
or
\be
ds^2=\frac{l^2}{\nu^2+3}\left[\cosh^2\rho d\xi^2+d\rho^2-\frac{4\nu^2}{\nu^2+3}(d\tau+\sinh\rho d\xi)^2\right],
\ee
it can be shown that they satisfy the equations of motion \eqref{e.o.m.metric} if 
\bea
&&\bar{\Lambda}_{0} = -\frac{1}{4\mu^{2}l^{4}} \left[4\mu^{2}l^{2}\bar{\sigma}\nu^{2}+12\mu l(\nu^{3}-\nu)+\gamma(-8\nu^{4}+18\nu^{2}-9)\right],\cr\nonumber\\&&
\tilde{\mu} l+3\nu= (\nu^2-1)\frac{\gamma}{\mu l},
\eea
which can be solved  to find $\nu$ and $l$ it terms of the parameters of the model.

For $\nu=1$ where $\tilde{\mu}l=-3$, the equations of motion admit a null warped AdS solution as follows
\bea
ds^{2} = l^{2}\left(\frac{dr^{2}}{r^{2}} +\frac{dudv}{r^{2}} + \frac{d^{2}u}{r^{4}} \right),\;\;\;\;\;\;\;
{\rm with}\;\;l^2= -\frac{1}{2\bar{\Lambda}_0}\left(\bar{\sigma}+\frac{\tilde{\mu}}{\mu}\right) 
\eea
%In this case the parameters of model are given by 
%\bea
%\bar{\Lambda}_{0} = -\frac{1}{4\mu^{2}l^{4}} (4\mu^{2}l^{2}\bar{\sigma}+\gamma),~~~~~~~\gamma = %-2\mu l (\mu l \bar{\sigma}+3),
%\eea
which is indeed the Schr\"odinger solution we have found in the previous section with the dynamical scaling
$z=2$.
% which corresponds to  

Note that in the $\gamma\rightarrow 0$ limit, the above solution reduces to that of TMG and, in particular we
get $3\nu=-\mu l$.  One may also consider a limit in which  the coefficient of the Cotton  
tensor vanishes, though the coefficient of  the curvature-squared symmetric tensor $J$ 
survives in this limit. This limit might be taken as $\mu\rightarrow \infty, \gamma\rightarrow \infty$ while keeping $\kappa=\frac{\gamma}{\mu^2}$ fixed. In this limit one still gets the warped AdS solution with
 \be\label{tt}
\bar{\Lambda}_0=\frac{\bar{\sigma}^2}{\kappa},\;\;\;\;\;\;\;\;l^2=\left(\nu^2-\frac{3}{2}\right)\frac{\bar{\sigma}}
{\bar{\Lambda}_0}.
\ee 
It is easy to show that the warped AdS geometry with the above parameters  solves the  
following equations of motion
\bea\label{new}
\bar{\sigma} G_{\mu\nu} +\bar{\Lambda}_{0} g_{\mu\nu} +\kappa J_{\mu\nu}=0.
\eea
which can be obtained  from the equations of motion of the MMG model by setting the  coefficient of the Cotton tensor to zero. One could also  show that the above equations 
of motion admit several vacuum solutions including AdS, warped AdS, null warped AdS and AdS-wave solutions. It is then natural to wonder whether this equation could also define a consistent 
three dimensional gravity. We note, however, that to have a consistent theory 
the Bianchi identity should also be satisfied \footnote{We would like to thank the referee for his/her comment on this point.}.

Actually from the equation \eqref{new}  and taking into account that the Bianchi identity is satisfied for the Einstein tensor one gets $D_\mu J^{\mu\nu}=0$.  On the other hand  from a 
direct computation one has\cite{Bergshoeff:2014pca}
\be
\sqrt{-g}\;D_\mu J^{\mu\nu}=\epsilon^{\nu\rho\sigma} S_\rho^\tau C_{\sigma\tau},\;\;\;\;\;\;\;\;\;\;
{\rm with}\;\;\;S_{\mu\nu}=R_{\mu\nu}-\frac{1}{4}g_{\mu\nu} R.
\ee 
To make the theory consistent one should set the right hand side of the above equation to zero
\be\label{cons}
\epsilon^{\nu\rho\sigma} S_\rho^\tau C_{\sigma\tau}=0,
\ee
for all $\nu$. Therefore the equation \eqref{new} may define a consistent three dimensional
gravity if the  above constraint holds. This constraint guarantees  that the Bianchi identity is
satisfied on shell. It is easy to see that the constraint \eqref{cons} is satisfied by 
the  warped AdS solution \eqref{tt}.

To find further consistent solutions of the
model defined by the equation \eqref{new} consider the following
ansatz for a black hole solution 
\be
ds^2=\frac{l^2}{r^2}\left(-f(r) dt^2+\frac{dr^2}{f(r)}+dx^2\right).
\ee
Plugging this ansatz into the constraint equation \eqref{cons} one 
finds
\be
f(r)=1+b_0 r+b_1 r^2.
\ee
where $b_0$ could be zero. Indeed when $b_0$ is zero the ansatz is a solution of 
the equation \eqref{new} if 
\be\label{ss}
\bar{\Lambda}_0 =-\frac{\kappa +4 l^2 \bar{\sigma} }{4 l^4}.
\ee 
On the other hand if one requires $b_0\neq 0$, then the above ansatz is a solution 
at a particular point\footnote{ This point was called {\it merger point} in
\cite{Arvanitakis:2014yja}}  in the moduli space of the  parameters of the model 
\be\label{ss1}
\bar{\Lambda}_0=-\frac{\bar{\sigma}}{2l^2},\;\;\;\;\;\;\;\;\;\;\;\kappa=-2l^2\bar{\sigma}.
\ee
Note that since the constraint \eqref{cons} is independent of the parameters of the model 
and taking into account that the Cotton tensor for these solutions is identically zero,
the above solutions are also solutions of the equations of motion of MMG. It is also straightforward to see
that in the merger  
point the equation \eqref{new} admits a null solutions, though in this case since
the Cotton tensor is non-zero the corresponding null solution cannot be a solution
of MMG. Of course as we have seen the MMG model has its own null solution,
though it occurs at another point in the moduli space of the parameters.

The model also admits the wave solution
\bea\label{wm}
ds^{2} = \frac{l^{2}}{r^{2}} \bigg(dr^{2}-2dudv - F(r) du^{2}\bigg),
\eea
 for which the constraint \eqref{cons} is satisfied identically. From the equation \eqref{new} one finds $F(r)=1 +c\; r^2$ if the parameters of the model
 satisfy the equation \eqref{ss}.  It is worth noting that 
 at the merger point \eqref{ss1} both the constraint  and the equation of motion \eqref{new}
are satisfied identically by the wave metric \eqref{wm}  for arbitrary function $F$. In other words
the function $F$ cannot be fixed by the equation of motion.  We should admit that 
the significant of this solution is not clear to us. It would be interesting  to explore this
point better. 
Note that this is not the case for MMG where at this point the function $F$ is fixed 
by the corresponding equations of motion to be $F=1+c_0 r+c_1 r^2$.

To conclude we note that the equation \eqref{new} together with the constraint \eqref{cons}
may define a consistent three dimensional gravity. Note that since the resulting mode has 
several non-trivial vacuum solutions besides  that of an  AdS, it should not be equivalent 
to three dimensional  Einstein gravity.

%%%%%%%%%%%%%%%%%%%%%%%%%%%%%%%%%%%%%%%%%%%%%%%%%%%%%%%%%%%%%%%%%
%%%%%%%%%%%%%%%%%%%%%%%%%%%%%%%%%%%%%%%%%%%%%%%%%%%%%%%%%%%%%%%%%

\section{Logarithmic CFT and new anomaly}

In sections  two and three  we have found that for particular values of the parameters of the MMG model, 
it can admit logarithmic solutions. In the case of the TMG model it was  conjectured\cite{Grumiller:2008qz}
 that when the logarithmic modes appear, the dual field theory would be   a LCFT with one zero central charge \cite{{Skenderis:2009nt},{Grumiller:2009mw}}.  For the NMG model \cite{Bergshoeff:2009hq}
the situation is the same though both the central charges are zero (see
for example \cite{{Grumiller:2009sn},{Alishahiha:2010bw}}). 

When the central charge is zero, the two point 
function  of the corresponding energy momentum tensor should vanish as well. We note, however, that
in a LCFT one has the logarithmic partner of stress-tensor  which could have non-zero
two point function whose expression is fixed by a new parameter  known
as ``new anomaly''.  Since the local structure of MMG is the same as TMG, one would expect to see
the same behavior for MMG too. Indeed this is the aim of this section to evaluate the 
new anomaly for MMG model.

To proceed let us recall that  the asymptotic symmetry algebra of AdS geometry in MMG model 
consists of two copies of the Virasoro algebra with central charges\cite{Bergshoeff:2014pca}
\bea\label{central.charge}
c_{\pm} = \frac{3l}{2G} \left( \sigma \pm \frac{1}{\mu l}+ \frac{\alpha-\alpha^2 \Lambda_{0}l^2}
{2\mu^{2}l^2(1+\sigma\alpha)^{2}} \right).
\eea
Note that in the TMG limit $\alpha \rightarrow 0$ the above central charges reduce to the 
that of  TMG. By making use of  the equation \eqref{Lambda0}, the expression for 
central charges may be recast into the following form
\be
c_{\pm} = \frac{3}{2\alpha\mu G} \big(\sqrt{\alpha^{2}+\mu ^{2}l^{2}(1+\sigma\alpha)^{2}}-\mu l\pm \alpha\big).
%c_\pm=\frac{3l}{2G}\left[2(\sigma\pm\frac{1}{\mu l})+\alpha\mp\frac{1}{\mu l}(1\pm \tilde{\mu}l)\right],
\ee
It is then straightforward to evaluate the value of the central charges at the  critical 
point $\tilde{\mu} l=-1$ where  the corresponding geometry could provide
 a holographic description for a LCFT
\bea
\tilde{\mu}l =-1\;: &&[{\rm or}\;\alpha = -2 ( \sigma -\frac{1}{\mu l})]~~~~~~~\rightarrow ~~~~~~~ 
c_{+} = \frac{3}{\mu G},~~~~~~c_{-} = 0,
%\tilde{\mu}l =1\; :&&\alpha = -{2} ( \sigma +\frac{1}{\mu l})~~~~~~~\rightarrow ~~~~~~~ 
%c_{+} = 0,~~~~~~~~~c_{-} = -\frac{3}{\mu G}.
\eea
As we see  the central charge of the left moving sector  is zero and therefore one 
  should naturally  consider  a new  mode corresponding to the logarithmic partner
of the energy momentum tensor whose two point correlation function would be fixed by the
new anomaly. 
Following \cite{Grumiller:2010tj, Grumiller:2013at} the new anomaly can be evaluated as follows
\bea\label{bL}
b_{L} \equiv b_{-} = \lim_{\alpha \rightarrow \alpha_{c}} \frac{c_{-}}{h^{(0)} -h^{(m)}}.
\eea
where $\alpha_{c}=-2 ( \sigma -\frac{1}{\mu l})$. Note that in this limit both the
denominator and the numerator vanish, though their ratio remains finite and nonzero. More precisely,  from \eqref{Bran} and using the identity \eqref{Id} one gets
\bea
h^{(0)} -h^{(m)} =\frac{1}{2\alpha} \big(\alpha-\mu l (1+\sigma\alpha)^{2}+\sqrt{\alpha^{2}+ \mu^{2}l^{2}(1+\sigma\alpha)^{2}}~\big).
\eea
Plugging this expression into the equation \eqref{bL} and performing the limit $\alpha\rightarrow \alpha_c$ one arrives at
\bea
b_{-} =-\frac{3l}{G}\frac{\mu l}{(\mu l \sigma-2)^{2}}.
\eea
Clearly in the TMG limit where $\sigma=1 ,\mu l =1$ one has
\bea
c_{+} = \frac{3l}{ G},~~~~~~~~~~~~~ b_{-} = -\frac{3l}{ G},~~~~~~~~~~~~~c_-=0,
\eea
in agreement with the results of \cite{Skenderis:2009nt,Grumiller:2009mw}.

Therefore denoting by $(T(z), \bar{T}(\bar z))$ and $(t(z),\bar{t}(\bar z))$ the energy momentum tensor and its logarithmic 
partner, respectively, it is natural to expect that the three dimensional  MMG gravity at the critical point would provide a holographic dual for a
LCFT such that
\bea
&&\langle T(z) T(0)\rangle =\frac{c_-}{2z^4}=0,\;\;\;\;\;\;\langle \bar{T}(\bar{z}) \bar{T}(0)\rangle =\frac{3l/G}{2\bar{z}^4}, \;\;\;\;\;\; \langle {T}(z) {t}(0)\rangle =\frac{\mu l}{(\mu l \sigma-2)^{2}}\frac{-3l/G}{2{z}^4},\nonumber \\
&&\langle {t}(z) {t}(0)\rangle =\frac{3l}{G}\frac{\mu l}{(\mu l \sigma-2)^{2}}\frac{\ln(m^2 z^2)}{{z}^4},
\eea
where $m$ is a scale. It would be interesting to derive the above equation using the holographic renormalization procedure (see for example \cite{Skenderis:2009nt} for TMG and \cite{Alishahiha:2010bw}
for NMG).

%%%%%%%%%%%%%%%%%%%%%%%%%%%%%%%%%%%%%%%%%%%%%%%%%%%%%%%%%%%%%%%%%%%%%%
%%%%%%%%%%%%%%%%%%%%%%%%%%%%%%%%%%%%%%%%%%%%%%%%%%%%%%%%%%%%%%%%%%%%%%

\section{Conclusions}

In this paper using the metric formulation of the  MMG model we have  studied 
linearized equations of motion  around an AdS vacuum solution. We have seen that 
the model has locally the same structure as that of  the TMG model. 
More precisely, in the linearized level,  upon a redefinition of the parameters, the corresponding equations of motion exactly reduce to that of TMG and therefore both models have the same spectrum. In particular there are
critical points at which the massive mode degenerates with the massless graviton. Of course there is big 
difference between MMG and TMG in the sense that although in  TMG one is forced to be at 
the critical points, a priori, there is no such requirement in MMG. In fact as it was shown in \cite{Bergshoeff:2014pca}
there is a wide range of the parameters where the model would be well defined.

Indeed, although the resulting equations for ther perturbations around an AdS vacuum
 are the same for both TMG and MMG models, the two models  have different conserved charges. This is indeed the reason why one could get a non-trivial
well defined 3D gravity from MMG model. Actually, since in the metric formulation of MMG one has no action, it is tricky to find the conserved charges of the model. 

We note, however, that even though one cannot 
write an action for MMG in the metric formulation, since in the first order of  perturbation,
 the resulting equations are exactly the same as that in TMG, one would expect 
 that, at least, at second order in perturbation it is possible to write an action whose equations of motion are the corresponding  linearized equations. Therefore it should be
possible to read off the energy of different modes using the results of TMG. In particular from  equations 
(70)-(72) of \cite{Li:2008dq}, up to numerical factors, one gets
\bea
E_M&\sim& \frac{M^2}{\mu G}\int d^3x \sqrt{-\bar{g}}\epsilon_{\beta}^{\;0\mu}h_{M}^{\beta\nu}\dot{h}_{M\mu\nu},\cr
E_L&\sim& -\frac{c_-}{l G}\int d^3x \sqrt{-\bar{g}}\bar{\nabla}^0 h_{L}^{\mu\nu}\dot{h}_{L\mu\nu},\cr
E_R&\sim& -\frac{c_+}{ l G}\int d^3x \sqrt{-\bar{g}}\bar{\nabla}^0 h_{R}^{\mu\nu}\dot{h}_{R\mu\nu}.
\eea
Here we have used a  decomposition by which   $h_{\mu\nu}=h_{M\mu\nu}+h_{L\mu\nu}+h_{R \mu\nu}$ and also 
$M^2$ and $c_{\pm}$ are given by \eqref{eqmass} and  \eqref{central.charge}, respectively.

We have also found AdS wave solutions of the model and shown that at the critical points 
the model  exhibit logarithmic solutions. Therefore at the critical point the model could provide a holographic dual 
for a LCFT. We have also calculated the new anomaly of the theory.

We have observed that in the metric formulation of MMG if one sets the coefficient of 
the Cotton tensor to zero, it  might still define a new consistent three dimensional gravity.
 Indeed a consistency condition requires to have the constraint 
 $\epsilon^{\nu\rho\sigma} S_\rho^\tau C_{\sigma\tau}=0$ as well.

It is important to mention that  since the resultant model has  several vacuum solutions 
besides  the AdS one, it should not be equivalent to three dimensional  Einstein gravity.
Even though it has a common feature with 3D conformal gravity, since the model has 
a null solution it cannot be equivalent to three dimensional conformal gravity either.
We note also that since the constraint \eqref{cons} is independent of the parameters appearing
in the equations of motion, any solution of the MMG model is a solution of the new
model defined by the equation \eqref{new}. Though the other way around is not 
correct. In particular at the merger point the new model has the wave solution \eqref{wm} 
with arbitrary $F$, though in MMG the function $F$ is fixed.

In the Chern-Simons like formalism of the MMG model  setting the coefficient of the Chern-Simons term to zero one arrives at\footnote{Note that to maintain $\gamma\rightarrow \infty$ one has 
to set $\sigma\alpha\rightarrow -1.$}
\be\label{uu}
L=-\sigma e\cdot R+\frac{\Lambda_0}{6} \; e\cdot e\times e+h\cdot T(\omega)+\frac{\alpha}{2}
e\cdot h\times h.
\ee
which as pointed out in \cite{Bergshoeff:2014pca} should be equivalent to the Einstein gravity.
It is, however, important to note that in general the equations of motion obtained from this action  
cannot be mapped to that defined by the 
equation \eqref{new}. Therefore there is no conflict between the results of our paper with 
the fact that the model  \eqref{uu} could be equivalent to the Einstein gravity.
% It would be interesting to explore this  point better and in particular to see
%if the model defined by \eqref{new} would lead to a consistent  three dimensional gravity.

To further explore the properties of MMG it would be interesting to study holographic
renormalization of the model. We note, however, that since in this context the action plays an essential role one will have to work with the Chern-Simons like formulation of the model.

{\bf Note added:} When we were in the final stage of submitting  our paper, the paper \cite{Tekin:2014jna} appeared
in the arXiv which has some overlap with ours.

\section*{Acknowledgements}
We would like to thank Hamid Afshar and Mahmoud safari  for useful discussions. M. M. Q.  also thanks Farhad Ardalan, Hessamaddin Arfaei and Shahin Rouhani for encouragement and support.

%---------------------------------------------------------------------
%Bibliography

%--------------------------------------------------------------------

\begin{thebibliography}{}

%\cite{Bergshoeff:2014pca}
\bibitem{Bergshoeff:2014pca} 
  E.~Bergshoeff, O.~Hohm, W.~Merbis, A.~J.~Routh and P.~K.~Townsend,
  ``Minimal Massive 3D Gravity,''
  Class.\ Quant.\ Grav.\  {\bf 31}, 145008 (2014)
  [arXiv:1404.2867 [hep-th]].
  %%CITATION = ARXIV:1404.2867;%%

%\cite{Deser:1981wh}
\bibitem{Deser:1981wh} 
  S.~Deser, R.~Jackiw and S.~Templeton,
  ``Topologically Massive Gauge Theories,''
  Annals Phys.\  {\bf 140}, 372 (1982)
  [Erratum-ibid.\  {\bf 185}, 406 (1988)]
  [Annals Phys.\  {\bf 185}, 406 (1988)]
  [Annals Phys.\  {\bf 281}, 409 (2000)].
  %%CITATION = APNYA,140,372;%%
  %2209 citations counted in INSPIRE as of 01 Sep 2014



%\cite{Hohm:2012vh}
\bibitem{Hohm:2012vh} 
  O.~Hohm, A.~Routh, P.~K.~Townsend and B.~Zhang,
  ``On the Hamiltonian form of 3D massive gravity,''
  Phys.\ Rev.\ D {\bf 86}, 084035 (2012)
  [arXiv:1208.0038].
  %%CITATION = ARXIV:1208.0038;%%
  %17 citations counted in INSPIRE as of 01 Sep 2014





%\cite{Moussa:2003fc}{Bouchareb:2007yx}{Anninos:2008fx}{Moussa:2008sj}
\bibitem{Moussa:2003fc} 
  K.~A.~Moussa, G.~Clement and C.~Leygnac,
  ``The Black holes of topologically massive gravity,''
  Class.\ Quant.\ Grav.\  {\bf 20}, L277 (2003)
  [gr-qc/0303042].
  %%CITATION = GR-QC/0303042;%%
  %84 citations counted in INSPIRE as of 03 Oct 2014

%\cite{Bouchareb:2007yx}{Anninos:2008fx}{Moussa:2008sj}
\bibitem{Bouchareb:2007yx} 
  A.~Bouchareb and G.~Clement,
  ``Black hole mass and angular momentum in topologically massive gravity,''
  Class.\ Quant.\ Grav.\  {\bf 24}, 5581 (2007)
  [arXiv:0706.0263 [gr-qc]].
  %%CITATION = ARXIV:0706.0263;%%
  %86 citations counted in INSPIRE as of 03 Oct 2014


%\cite{Anninos:2008fx}{Moussa:2008sj}
\bibitem{Anninos:2008fx} 
  D.~Anninos, W.~Li, M.~Padi, W.~Song and A.~Strominger,
  ``Warped AdS(3) Black Holes,''
  JHEP {\bf 0903}, 130 (2009)
  [arXiv:0807.3040 [hep-th]].
  %%CITATION = ARXIV:0807.3040;%%
  %173 citations counted in INSPIRE as of 01 Sep 2014



%\cite{Moussa:2008sj}
\bibitem{Moussa:2008sj} 
  K.~A.~Moussa, G.~Clement, H.~Guennoune and C.~Leygnac,
  ``Three-dimensional Chern-Simons black holes,''
  Phys.\ Rev.\ D {\bf 78}, 064065 (2008)
  [arXiv:0807.4241 [gr-qc]].
  %%CITATION = ARXIV:0807.4241;%%
  %45 citations counted in INSPIRE as of 03 Oct 2014





%\cite{Li:2008dq}
\bibitem{Li:2008dq} 
  W.~Li, W.~Song and A.~Strominger,
  ``Chiral Gravity in Three Dimensions,''
  JHEP {\bf 0804}, 082 (2008)
  [arXiv:0801.4566 [hep-th]].
  %%CITATION = ARXIV:0801.4566;%%
  
  %\cite{Grumiller:2008qz}
\bibitem{Grumiller:2008qz} 
  D.~Grumiller and N.~Johansson,
  ``Instability in cosmological topologically massive gravity at the chiral point,''
  JHEP {\bf 0807}, 134 (2008)
  [arXiv:0805.2610 [hep-th]].
  %%CITATION = ARXIV:0805.2610;%%
  %159 citations counted in INSPIRE as of 01 Sep 2014


%\cite{Alishahiha:2011yb}{Gullu:2011sj}{Bergshoeff:2011ri}
\bibitem{Alishahiha:2011yb} 
  M.~Alishahiha and R.~Fareghbal,
  ``D-Dimensional Log Gravity,''
  Phys.\ Rev.\ D {\bf 83}, 084052 (2011)
  [arXiv:1101.5891 [hep-th]].
  %%CITATION = ARXIV:1101.5891;%%
  %47 citations counted in INSPIRE as of 01 Sep 2014


%\cite{Gullu:2011sj}{Bergshoeff:2011ri}
\bibitem{Gullu:2011sj} 
  I.~Gullu, M.~Gurses, T.~C.~Sisman and B.~Tekin,
  ``AdS Waves as Exact Solutions to Quadratic Gravity,''
  Phys.\ Rev.\ D {\bf 83}, 084015 (2011)
  [arXiv:1102.1921 [hep-th]].
  %%CITATION = ARXIV:1102.1921;%%
  %35 citations counted in INSPIRE as of 01 Sep 2014
  

%\cite{Bergshoeff:2011ri}
\bibitem{Bergshoeff:2011ri} 
  E.~A.~Bergshoeff, O.~Hohm, J.~Rosseel and P.~K.~Townsend,
  ``Modes of Log Gravity,''
  Phys.\ Rev.\ D {\bf 83}, 104038 (2011)
  [arXiv:1102.4091 [hep-th]].
  %%CITATION = ARXIV:1102.4091;%%
  %54 citations counted in INSPIRE as of 01 Sep 2014
  

%\cite{Johansson:2012fs}
\bibitem{Johansson:2012fs} 
  N.~Johansson, A.~Naseh and T.~Zojer,
  ``Holographic two-point functions for 4d log-gravity,''
  JHEP {\bf 1209}, 114 (2012)
  [arXiv:1205.5804 [hep-th]].
  %%CITATION = ARXIV:1205.5804;%%
  


%\cite{AyonBeato:2004fq}
\bibitem{AyonBeato:2004fq} 
  E.~Ayon-Beato and M.~Hassaine,
  ``pp waves of conformal gravity with self-interacting source,''
  Annals Phys.\  {\bf 317}, 175 (2005)
  [hep-th/0409150].
  %%CITATION = HEP-TH/0409150;%%
  %41 citations counted in INSPIRE as of 10 Nov 2014



%\cite{AyonBeato:2005qq}
\bibitem{AyonBeato:2005qq} 
  E.~Ayon-Beato and M.~Hassaine,
  ``Exploring AdS waves via nonminimal coupling,''
  Phys.\ Rev.\ D {\bf 73}, 104001 (2006)
  [hep-th/0512074].
  %%CITATION = HEP-TH/0512074;%%
  %38 citations counted in INSPIRE as of 10 Nov 2014












%\cite{Clement:2009ka}
\bibitem{Clement:2009ka} 
  G.~Clement,
  ``Black holes with a null Killing vector in new massive gravity in three dimensions,''
  Class.\ Quant.\ Grav.\  {\bf 26}, 165002 (2009)
  [arXiv:0905.0553 [hep-th]].
  %%CITATION = ARXIV:0905.0553;%%
  %40 citations counted in INSPIRE as of 03 Oct 2014


%\cite{Clement:2009gq}
\bibitem{Clement:2009gq} 
  G.~Clement,
  ``Warped AdS(3) black holes in new massive gravity,''
  Class.\ Quant.\ Grav.\  {\bf 26}, 105015 (2009)
  [arXiv:0902.4634 [hep-th]].
  %%CITATION = ARXIV:0902.4634;%%
  %85 citations counted in INSPIRE as of 03 Oct 2014

%\cite{Arvanitakis:2014yja}
\bibitem{Arvanitakis:2014yja} 
  A.~S.~Arvanitakis, A.~J.~Routh and P.~K.~Townsend,
  ``Matter coupling in 3D "Minimal Massive Gravity",''
  arXiv:1407.1264 [hep-th].
  %%CITATION = ARXIV:1407.1264;%%
  %2 citations counted in INSPIRE as of 09 Sep 2014






 %\cite{Skenderis:2009nt}{Grumiller:2009mw}
\bibitem{Skenderis:2009nt} 
  K.~Skenderis, M.~Taylor and B.~C.~van Rees,
  ``Topologically Massive Gravity and the AdS/CFT Correspondence,''
  JHEP {\bf 0909}, 045 (2009)
  [arXiv:0906.4926 [hep-th]].
  %%CITATION = ARXIV:0906.4926;%%
    
  %\cite{Grumiller:2009mw}
\bibitem{Grumiller:2009mw} 
  D.~Grumiller and I.~Sachs,
  ``AdS (3) / LCFT (2)  $\rightarrow$  Correlators in Cosmological Topologically Massive Gravity,''
  JHEP {\bf 1003}, 012 (2010)
  [arXiv:0910.5241 [hep-th]].
  %%CITATION = ARXIV:0910.5241;%%
   
%\cite{Bergshoeff:2009hq}
\bibitem{Bergshoeff:2009hq} 
  E.~A.~Bergshoeff, O.~Hohm and P.~K.~Townsend,
  %``Massive Gravity in Three Dimensions,''
  Phys.\ Rev.\ Lett.\  {\bf 102}, 201301 (2009)
  [arXiv:0901.1766 [hep-th]].
  %%CITATION = ARXIV:0901.1766;%%
  %338 citations counted in INSPIRE as of 04 Sep 2014




%\cite{Grumiller:2009sn}{Alishahiha:2010bw}
\bibitem{Grumiller:2009sn} 
  D.~Grumiller and O.~Hohm,
  ``AdS(3)/LCFT(2): Correlators in New Massive Gravity,''
  Phys.\ Lett.\ B {\bf 686}, 264 (2010)
  [arXiv:0911.4274 [hep-th]].
  %%CITATION = ARXIV:0911.4274;%%
  %58 citations counted in INSPIRE as of 04 Sep 2014






%\cite{Alishahiha:2010bw}
\bibitem{Alishahiha:2010bw} 
  M.~Alishahiha and A.~Naseh,
  ``Holographic renormalization of new massive gravity,''
  Phys.\ Rev.\ D {\bf 82}, 104043 (2010)
  [arXiv:1005.1544 [hep-th]].
  %%CITATION = ARXIV:1005.1544;%%
  %33 citations counted in INSPIRE as of 04 Sep 2014





%\cite{Kwon:2011ey}
\bibitem{Kwon:2011ey} 
  Y.~Kwon, S.~Nam, J.~D.~Park and S.~H.~Yi,
  ``Quasi Normal Modes for New Type Black Holes in New Massive Gravity,''
  Class.\ Quant.\ Grav.\  {\bf 28}, 145006 (2011)
  [arXiv:1102.0138 [hep-th]].
  %%CITATION = ARXIV:1102.0138;%%
  %17 citations counted in INSPIRE as of 08 Sep 2014


  
%\cite{Grumiller:2010tj}
\bibitem{Grumiller:2010tj} 
  D.~Grumiller, N.~Johansson and T.~Zojer,
  ``Short-cut to new anomalies in gravity duals to logarithmic conformal field theories,''
  JHEP {\bf 1101}, 090 (2011)
  [arXiv:1010.4449 [hep-th]].
  %%CITATION = ARXIV:1010.4449;%% 
  
  
%\cite{Grumiller:2013at}
\bibitem{Grumiller:2013at} 
  D.~Grumiller, W.~Riedler, J.~Rosseel and T.~Zojer,
  ``Holographic applications of logarithmic conformal field theories,''
  J.\ Phys.\ A {\bf 46}, 494002 (2013)
  [arXiv:1302.0280 [hep-th]].
  %%CITATION = ARXIV:1302.0280;%%  
  

%\cite{Tekin:2014jna}
\bibitem{Tekin:2014jna} 
  B.~Tekin,
  ``Minimal Massive Gravity: Conserved Charges, Excitations and the Chiral Gravity Limit,''
  arXiv:1409.5358 [hep-th].
  %%CITATION = ARXIV:1409.5358;%%

 
\end{thebibliography}
\end{document}